\documentclass{ws-procs9x6}

\usepackage{amsmath,amssymb,graphicx,slashbox}
\usepackage{cite}

\def\mydate{April 6, 2007}
\def\ignore#1{{}}

\newcommand{\beeq}{\begin{equation}}
\newcommand{\eneq}{\end{equation}}
\newcommand{\beqn}{\begin{eqnarray}}
\newcommand{\eeqn}{\end{eqnarray}}

\def\mybig{\displaystyle \strut }

\def\dd{\partial}
\def\la{\raise.16ex\hbox{$\langle$}\lower.16ex\hbox{}  }
\def\ra{\, \raise.16ex\hbox{$\rangle$}\lower.16ex\hbox{} }
\def\go{\rightarrow}

\def\onehalf{ \hbox{${1\over 2}$} }
\def\half{ {1\over 2} }

\def\vphi{\varphi}
\def\ep{\epsilon}
\def\psibar{ \psi \kern-.65em\raise.6em\hbox{$-$} }

\def\eff{{\rm eff}}

\def\Pl{{\rm Pl}}

\def\to{$^{\hbox{-}}$}
\def\mycite#1#2{\cite{#1}\to\cite{#2}}

\def\myfrac#1#2{{\mybig #1\over \mybig #2}}

\def\cG{{\cal G}}

\begin{document}

\noindent
{\small \mydate \hfill OU-HET 579/2007}
\vskip -.5cm

\title{Gauge-Higgs Unification and LHC/ILC}

\author{Yutaka Hosotani$^*$}

\address{Department of Physics, Osaka University,\\
Toyonaka, Osaka 560-0043, Japan\\
$^*$E-mail: hosotani@phys.sci.osaka-u.ac.jp}

\begin{abstract}
In the gauge-Higgs unification scenario the 4D Higgs field is identified with 
the zero mode of the extra-dimensional component of gauge potentials.
The mass of the Higgs particle in the unification in the Randall-Sundrum 
warped spacetime is predicted to be 
in the range 100 GeV - 300 GeV.  The $WWZ$ gauge couplings  
remains almost universal as in the standard model, but substantial deviation
results for the Higgs couplings.    The  $WWH$ and $ZZH$ couplings
are suppressed by a factor $\cos \theta_H$ from the values in the 
standard model, where $\theta_H$ is the
Yang-Mills AB phase along the fifth dimension.  
These can be tested at LHC and ILC.
\end{abstract}


\bodymatter

\section{Origin of the Higgs boson}

There is one particle missing in the standard model of electroweak 
interactions.  It is the Higgs boson.  The Higgs boson must exist, 
either as an elementary particle or as a composite 
particle.  The electroweak unification is possible, only if there is
something which breaks  $SU(2)_L \times U(1)_Y$ symmetry
to $U(1)_{EM}$ symmetry.  In the standard model   the Higgs
boson,  whose potential is such that the electroweak symmetry 
is spontaneously broken, gives masses to $W$ and $Z$ bosons.
It also gives quarks and leptons masses through Yukawa couplings.

The standard model  seems economical, but it  hides dirty
secret. Physics ought to be based on simple principles, but there 
seems no good principle for the Higgs sector.  As a result
the standard model is afflicted with many arbitrary parameters.
There have been many proposals.  Technicolor theory views the 
Higgs boson as a composite state resulting from strong
technicolor interactions.  Supersymmetry (SUSY) is a leading
candidate beyond the standard model which cures the gauge
hierarchy problem. However, the situation concerning a large 
number of arbitrary parameters becomes worse in the minimal 
supersymmetric standard model.
There are other proposals such as the little Higgs theory and 
the Higgsless theory as well.

In this article  I would like to argue that the Higgs field is ``clean''. 
The Higgs field is a part of gauge fields in higher dimensions, 
the Higgs sector being controlled  by the gauge principle.
The difference between the Higgs particle and gauge bosons 
originates from the structure of the extra-dimensional space.
The scenario is called the gauge-Higgs unification.

The gauge-Higgs unification scenario can
be tested at LHC and ILC.\cite{HM}$^{\hbox{-}}$\cite{HS2}  
It predicts that the mass of the Higgs 
particle is around 100 GeV - 300 GeV, exactly in the energy
region where LHC can explore.  Further the couplings of the 
Higgs particle to the W and Z bosons, and also to quarks and 
leptons are substantially reduced compared with those in 
the standard model.  Thus the Higgs experiments at LHC may
uncover the origin of the Higgs particle, and disclose 
the existence of extra dimensions.

\section{Old gauge-Higgs unification}
The idea of the gauge-Higgs unification is very old.\cite{Fairlie1, Manton2, Manton1}
In the Kaluza-Klein
theory the gravity in five dimensional spacetime of topology
$M^4 \times S^1$ unifies the four-dimensional gravity with the 
electromagnetism.  The part of the metric, $g_{\mu 5}$ ($\mu=0,1,2,3$) ,
contains the 4D vector potential $A_\mu$ in the electromagnetism.
In the gauge-Higgs unification one considers gauge theory, instead of
gravity, in higher dimensional spacetime.  Extra-dimensional 
components, $A_{y_j}$, of gauge potentials transform as 4D scalars under 4D
Lorentz transformations.   The 4D Higgs field is identified with a 
low energy mode of $A_{y_j}$.  The Higgs field becomes a part of gauge
fields.   

This scenario was proposed by Fairlie and by Forgacs and Manton 
in 1979.
  They tried to achieve unification by restricting configurations of
gauge fields in extra dimensions with symmetry ansatz.    
In ref.\ 7  \ignore{\cite{Manton1}}
Manton considered gauge theory with gauge group $\cG$ 
defined on $M^4 \times S^2$.   It is assumed that only spherically symmetric
configurations are allowed and gauge fields have 
non-vanishing flux (field strengths) on $S^2$.  Further it is demanded 
that the gauge group $\cG$ breaks down to $SU(2)_L \times U(1)_Y$ 
by non-vanishing flux.  There appears a Higgs doublet as a low energy mode
of $A_{y_j}$.  Quite amazingly the Higgs doublet turns out to have a negative
mass squared so that the symmetry further breaks down to $U(1)_{EM}$.

There are two parameters; the radius $R$ of $S^2$ and the gauge 
coupling $g_6$ in the six-dimensional spacetime.  These two parameters 
are fixed by the Fermi constant and the four-dimensional $SU(2)_L$
gauge coupling $g$.    $m_W$, $m_Z$, and $m_H$ are determined
as functions of $g_6$ and $R$.  The Weinberg angle $\theta_W$ is
determined by the gauge group only.  There are  three gauge groups
which satisfy the above requirements.  The result is summarized in 
Table \ref{manton-table}.

\begin{table}[t]
\tbl{Spectrum in the gauge-Higgs unification model\cite{Manton1} by Manton.}
{\begin{tabular}{|c|c|c|c|c|}
\hline \rule[-2mm]{0mm}{5mm} 
 $\cG$ & $\sin^2 \theta_W$ & 
 $m_W$ & $m_Z$ & $m_H$  \\ 
 \hline  \rule[-1.8mm]{0mm}{5mm}
$SU(3)$& 3/4 & 44 GeV & 88 GeV & 88 GeV   \\ 
\hline \rule[-1.8mm]{0mm}{5mm}
$O(5)$& 1/2 & 54 GeV & 76 GeV & 76 GeV   \\ 
\hline \rule[-1.7mm]{0mm}{5mm}
$G_2$& 1/4 & 76 GeV & 88 GeV & 88 GeV   \\ \hline 
 \end{tabular}}
\label{manton-table}
\end{table}

The unification is achieved and the Higgs mass is predicted,
though numerical values are not realistic.
There are  generic problems in this scheme.  
First, the mass $m_Z$ is $\sim 1/R$.  In other words, it necessarily predicts  a too
small  Kaluza-Klein scale  $1/R$.  Secondly, and more importantly, there is no 
justification for the ansatz of non-vanishing flux.   The restriction to spherically 
symmetric configurations is not justified either.

\section{New gauge-Higgs unification}

There is a better way of achieving gauge-Higgs unification.  The key is
to consider gauge theory in a non-simply connected spacetime. 
It utilizes the Hosotani mechanism.\cite{YH1, YH2, HHHK, Hebecker}

\subsection{Yang-Mills AB phase $\theta_H$}

When the space is not simply connected, a configuration of vanishing 
field strengths  does not necessarily mean trivial.  The phenomenon is called
the Aharonov-Bohm (AB) effect in quantum mechanics.  Consider 
$SU(N)$ gauge theory on $M^4 \times S^1$ with coordinates
$(x^\mu, y)$, and impose periodic boundary conditions 
$A_M(x, y+ 2\pi R) = A_M(x,y)$.    A  configuration 
$A_y(x,y) =\,$constant gives $F_{MN}=0$, but gives
\beeq
W \equiv P \exp \bigg\{ ig \int_0^{2\pi R} dy \, A_y \bigg\} 
=  U \begin{pmatrix}
e^{i\theta_1} \cr & \ddots \cr && e^{i\theta_N}
\end{pmatrix} U^\dagger
\label{AB1}
 \eneq
where $U^\dagger = U^{-1}$ and $\sum_{j=1}^N \theta_J = 0$ ($mod~2\pi$). 
$\theta_j$'s are Yang-Mills AB phases in the theory, denoted  collectively
as $\theta_H$.
They cannot be eliminated by gauge transformations preserving the 
boundary conditions.   

Classical vacua are degenerate.  Yang-Mills AB phases $\theta_H$ 
label flat directions of the classical potential.  The degeneracy  is
lifted at the quantum level.  The mass spectrum $\{ m_n \}$ of various fields
depends on $\theta_H$.  The effective potential $V_\eff (\theta_H)$ 
is given at the one loop level  by
\beeq
V_\eff (\theta_H) = \sum \mp \frac{i}{2} \int
\frac{d^4 p}{(2\pi)^4} \sum_n \ln 
\big\{ -p^2 + m_n^2 (\theta_H) \big\} ~~.
\label{Veff1}
\eneq
The value of $\theta_H$ is determined by the location of the global
minimum of $V_\eff (\theta_H)$.

\subsection{Dynamical gauge symmetry breaking}

Once the matter content is specified, the effective potential is determined
and so is the value of $\theta_H$ in the true vacuum. 
Suppose that all fields are periodic so that the boundary conditions are $SU(N)$ 
symmetric.  If $\theta_H \not= 0$,  the symmetry breaks down to
a subgroup of $SU(N)$ in general.  In other words we have dynamical
gauge symmetry breaking.

Take $SU(3)$ as an example.  In a pure gauge theory the global minima
are located at 
$\theta_1=\theta_2= \theta_3 = 0, \frac{2}{3} \pi, \frac{4}{3} \pi$.
The $SU(3)$ symmetry is unbroken.  Add periodic fermions in the fundamental
representation. Then the global minimum is given by 
$\theta_1=\theta_2= \theta_3 = 0$, the $SU(3)$ symmetry remaining unbroken.
If one has, instead, a periodic fermion in the addjoint representation,
then the global minima are found at
$(\theta_1, \theta_2, \theta_3) = (0, \frac{2}{3} \pi, -\frac{2}{3} \pi)$ and
its permutations. The $SU(3)$ symmetry breaks down to $U(1) \times U(1)$.
These results are tabulated in Table \ref{symmetry-table}.
Dynamical gauge symmetry breaking occurs quite naturally.  It involves 
no fine tuning.\cite{YHscgt2}

\begin{table}[b]
\tbl{Dynamical gauge symmetry breaking in $SU(3)$ theory on $M^4 \times S^1$.
$N^F_{\rm fund}$ and $N^F_{\rm add}$ denote the number of periodic fermions
in the fundamental and addjoint representations, respectively.}
{\begin{tabular}{|c|c|c|}
\hline \rule[-2mm]{0mm}{5mm} 
 $(N^F_{\rm fund}, N^F_{\rm add})$ & global minima 
 $(\theta_1^{\rm min}, \theta_2^{\rm min}, \theta_3^{\rm min})$ & 
residual symmetry  \\ 
 \hline  \rule[-2mm]{0mm}{5mm}
$(~0~,~0~)$
& $(0,0,0), (\pm \frac{2}{3} \pi, \pm \frac{2}{3} \pi, \pm \frac{2}{3} \pi)$
&$SU(3)$   \\ 
\hline  \rule[-2mm]{0mm}{5mm}
$(~n~,~0~)$
& $(0,0,0)$
&$SU(3)$   \\ 
\hline  \rule[-2mm]{0mm}{5mm}
$(~0~,~n~)$
& $(0, + \frac{2}{3} \pi, - \frac{2}{3} \pi)$ + permutations
&$U(1) \times U(1)$   \\ 
\hline  \rule[-2mm]{0mm}{5mm}
$(~1~,~1~)$
& $(0,\pi,\pi)$  + permutations
&$SU(2) \times U(1)$     \\ \hline 
 \end{tabular}}
\label{symmetry-table}
\end{table}

Instead of periodic boundary conditions, one might impose more
general twisted boundary conditions. For instance, one can impose
$A_M(x, y+ 2\pi R) = \Omega A_M(x, y) \Omega^\dagger$ 
($\Omega \in SU(N)$).  It can be shown that on $M^4 \times S^1$ 
physics does not depend on the choice of $\Omega$, thanks to dynamics
of  Yang-Mills AB phases $\theta_H$.  On orbifolds such as $M^4 \times (S^1/Z_2)$
and the Randall-Sundrum warped spacetime there appear a finite number of 
inequivalent sets of boundary conditions.\cite{YH2, HHK, YH4}

\subsection{Finiteness of $V_\eff (\theta_H)$ and the Higgs mass}

A mode of four-dimensional fluctuations of Yang-Mills AB phase $\theta_H$
is identified with the 4D Higgs field in an appropriate setup.   Hence
$V_\eff (\theta_H)$ is directly related to the effective potential for the 4D
Higgs field $\vphi_H(x)$.

One significant feature is that the $\theta_H$-dependent part of 
$V_\eff (\theta_H)$ is finite.  The mass squared of the Higgs boson, $m_H^2$,
is essentially the curvature of $V_\eff (\theta_H)$ at its global minimum,
implying the finiteness of  $m_H^2$.\cite{Lim2}

The finiteness of $V_\eff (\theta_H)$ at the one loop level has been shown 
explicitly in various models.\cite{YH1,YH2}  
A general proof goes as follows. \cite{YHscgt2, YHfinite}

First of all   large gauge invariance in theory guarantees that
$\theta_H$ is related to $\theta_H + 2\pi$  by a large 
gauge transformation which preserves the boundary conditions.  It implies that
$V_\eff (\theta_H + 2\pi)=V_\eff (\theta_H)$  to all order
in perturbation theory.   $V_\eff (\theta_H)$ can be expanded in a 
Fourier series; $V_\eff (\theta_H) = \sum_n a_n^V  e^{in\theta_H}$.

The one loop effective potential is given by (\ref{Veff1}).  In flat space $S^1$
$m_n (\theta_H) = (n + \ell \theta_H/2\pi + \alpha) m_{KK}$.   Here
the Kaluza-Klein mass scale $m_{KK} = 1/R$ and $\ell$ is an integer.
$\alpha$ is a constant determined by the boundary condition of each field.
It follows that 
$V_\eff^{(k)} (\theta_H)= \dd^k V_\eff (\theta_H) / \dd \theta_H^k$ 
becomes finite for sufficiently large $k$ almost everywhere in $\theta_H$.  
$V_\eff^{(k)} (\theta_H)$ can develop infrared divergence at a discrete set
of values of $\theta_H$ where    $m_n(\theta_H)$ vanishes, namely
at a set of points of measure zero.  Hence
$n^k a_n^V$ ($n \not= 0$) becomes finite, implying the finiteness of
$V_\eff (\theta_H)$ at the one loop level.
The argument remains valid in the Randall-Sundrum warped spacetime 
as $m_n (\theta_H) \sim (n + \ell \theta_H/2\pi + \alpha) m_{KK}$
for $|n| \gg 1$.

The finiteness seems to hold beyond one loop.  It has been shown that
$m_H^2$in QED in $M^4 \times S^1$ is
finite  at the two loop level after renormalization in $M^5$.\cite{Maru1} 
There is nonperturbative lattice simulation indicating the finiteness 
as well.\cite{Irges}

\section{Electroweak interactions}

To apply  gauge-Higgs unification scenario to electroweak interactions,
several  features have to be taken into 
account.\cite{Pomarol1}${}^{\hbox{-}}$\cite{Lim3}
First, the electroweak symmetry is
$SU(2)_L \times U(1)_Y$, which breaks down to $U(1)_{EM}$.  The Higgs field
$\vphi_H$  is an $SU(2)_L$ doublet.  
In the gauge-Higgs unification the Higgs field is a part of 
gauge fields, or must belong to the adjoint representation of the gauge group 
$\cG$.   This means that $\cG$ must be larger than $SU(2)_L \times U(1)_Y$,
as Fairlie, Forgacs, and Manton originally pointed out.\cite{Fairlie1, Manton2, Manton1}

Second, fermion content is chiral.  This is highly nontrivial in higher dimensional
gauge theory, as a spinor in higher dimensions always contains both right- and 
left-handed components in four dimensions.  The left-right asymmetry in 
fermion modes at low energies can be induced from nontrivial topology of 
extra-dimensional space and non-vanishing flux of gauge fields in extra 
dimensions.\ignore{\cite{YH3}}
There is another, simpler and more powerful,  way to have chiral fermions.  
If the extra-dimensional space  is an orbifold, 
appropriate boundary conditions naturally give rise to chiral fermion
content.\cite{Pomarol1, Antoniadis1}

Let us illustrate how the orbifold structure fits in the gauge-Higgs unification,
by taking  gauge theory on $M^4 \times (S^1/Z_2)$.   
The orbifold $M^4 \times (S^1/Z_2)$
is obtained from $M^4 \times S^1$ by identifying $(x^\mu, -y)$ and $(x^\mu, y)$.
There appear two fixed points (branes) at $y=0$ and $y= \pi R$.
We define gauge theory on a covering space of $M^4 \times (S^1/Z_2)$, namely
for $-\infty < y < + \infty$, and impose restrictions such that physics is the 
same at $(x^\mu, y), (x^\mu, y+2\pi R)$ and $(x^\mu, -y)$.  The single-valuedness
of physics does not necessarily mean that vector potentials $A_M$ are single-valued.
In gauge theory they may be twisted by global gauge transformation.
More explicitly
\beeq
\begin{pmatrix}  A_\mu \cr A_y \end{pmatrix} (x, y_j -y) = 
P_j \begin{pmatrix}  A_\mu \cr-  A_y \end{pmatrix} (x, y_j +y)  P_j^{-1}
\quad (j=0, 1)
\label{BC1}
\eneq
where $y_0=0$ and $y_1=\pi R$.  Here $P_j$ is an element of the gauge 
group $\cG$ satisfying  ${P_j}^2 = 1$.   When $P_j \not\propto 1$, 
the gauge symmetry is partially broken by the boundary conditions.
The physical symmetry, in general,  can be different from 
the residual symmetry given by $(P_0 ,  P_1)$.  It can be either reduced
or enhanced by dynamics of $\theta_H$.\cite{HHHK}

To see how an $SU(2)$ doublet Higgs field emerges, take $\cG = SU(3)$
and $P_0= P_1 = {\rm diag} (-1,-1,1)$.  Then, the orbifold boundary condition
(\ref{BC1}) implies that $SU(2) \times U(1)$ part of the four-dimensional 
components $A_\mu$ are even under parity at $y=0, \pi R$, which
contains zero modes corresponding to $SU(2) \times U(1)$ gauge fields
in four dimensions.  On the other hand the extra-dimensional component 
$A_y$ has zero modes in the off-diagonal part;
\beeq
SU(3): ~ 
A_y = 
\begin{pmatrix} &&\phi^+ \cr && \phi^0 \cr 
\phi^{+*}& \phi^{0*} \end{pmatrix} ~~,~~
\Phi =  \begin{pmatrix} \phi^+ \cr \phi^0 \end{pmatrix}~.
\label{su3zero}
\eneq
The zero mode $\Phi$ becomes an $SU(2)$ doublet
Higgs field.  Take $\cG = SO(5)$ and
$P_0= P_1 = {\rm diag} (-1,-1,-1,-1, 1)$ as another example.   In this case
the  $SO(5)$ symmetry breaks down to 
$SO(4) \simeq SU(2)_L \times SU(2)_R$.  Zero modes of $A_y$ are
\beeq
SO(5): ~ 
A_y = 
\begin{pmatrix} 
&& && \phi_1\cr && && \phi_2\cr && && \phi_3\cr && && \phi_4\cr 
-\phi_1 &-\phi_2 &-\phi_3 &-\phi_4  \end{pmatrix} 
~~,~~
\Phi = \begin{pmatrix} \phi_1+ i\phi_2 \cr \phi_4 - i \phi_3 \end{pmatrix} ~.
\label{so5zero}
\eneq
$\Phi $ is an $SU(2)_L$ doublet.  $\Phi$ is related to the Yang-Mills AB phases 
by (\ref{AB1}).

Chiral fermions naturally emerge.  Take $\cG = SU(3)$ with $P_j$ in
(\ref{su3zero}).    Fermions in the fundamental representation of $SU(3)$
obey the boundary condition 
$\psi(x, y_j - y) = P_j \gamma^5 \psi (x, y_j + y)$ so that
\ignore{
\beeq
\psi(x, y_j - y) = P_j \gamma^5 \psi (x, y_j + y) ~~.
\label{BC2}
\eneq
}
$\psi $ is decomposed as
\beeq
\psi = \begin{pmatrix}
~\nu_L~ & ~\tilde \nu_R~ \cr
e_L & \tilde e_R \cr
\tilde e_L & e_R  \end{pmatrix} ~.
\label{fermion1}
\eneq
$\nu_L$, $e_L$ and $e_R$ have zero modes, whereas 
$\tilde \nu_R$, $\tilde e_R$ and $\tilde e_L$ do not.
Fermion content at low energies is chiral as desired.

\begin{figure}[h,t,b]
\centering  \leavevmode
\includegraphics[width=9cm]{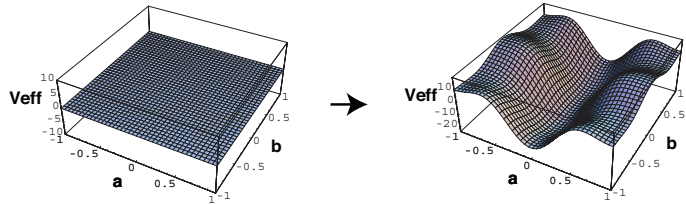}
\hskip .2cm
\caption{The effective potential $V_\eff(\theta_1, \theta_2)$ in the
$U(3) \times U(3)$ model in ref.\ 25 which has two
$\theta_H$'s,  $\theta_1=\pi a$ and 
$\theta_2 = \pi b$.  $V_\eff=0$ at the classical level (in the left figure),
but becomes nontrivial at the one loop level (in the right figure).}
\label{effV_flat}
\end{figure}

In the gauge-Higgs unification scenario the Higgs boson is massless
at the tree level.  Its mass is generated by radiative corrections.
The mass of the Higgs boson is determined by the curvature of 
the effective potential $V_\eff (\theta_H)$ at the minimum.
In fig.\ \ref{effV_flat} $V_\eff (\theta_1, \theta_2)$ is
displayed in the $U(3) \times U(3)$ model of ref.\ 25.

\section{Difficulties in flat spacetime}

The gauge-Higgs unification scenario in flat spacetime is afflicted with
a few intrinsic difficulties.  The electroweak symmetry is spontaneously broken
by $\theta_H$.  Non-vanishing $\theta_H$ gives rise to non-vanishing
masses for $W$ and $Z$ bosons.  $m_W$, for instance, is typically given by
\beeq
m_W \sim \frac{\theta_H}{2\pi} \times \frac{1}{R} \sim
 \frac{\theta_H}{2\pi} \times  m_{KK} ~~.
\label{mW1}
\eneq
Here $R$ is the size of the extra-dimensions.
Secondly, the effective potential $V_\eff(\theta_H)$ is generated at the one-loop
level, and therefore is $O(\alpha_W)$ where $\alpha_W= g_W^2/4\pi$ is the 
$SU(2)_L$ coupling.  The Higgs mass $m_H^2$ becomes $O(\alpha_W)$ as well.
Evaluation of $V_\eff$ shows that
\beeq
m_H \sim \sqrt{\alpha_W} \times  \frac{1}{R} \sim
\sqrt{\alpha_W}  ~  \frac{2\pi}{\theta_H}  ~ m_W ~~.
\label{mH1}
\eneq

The relations (\ref{mW1}) and (\ref{mH1}) are generic predictions from the 
gauge-Higgs unification in flat spacetime.  Once the value of $\theta_H$ is given,
$m_{KK}$ and $m_H$ are predicted.    The value of $\theta_H$ is determined from
the location of the global minimum of $V_\eff(\theta_H)$.  It depends on the matter
content in the theory.  Given standard matter content of quarks and leptons
with a minimal set of additional matter, the global minimum of $V_\eff(\theta_H)$
is typically located either at $\theta_H=0$ or at $\theta_H=(.2 \sim .8) \pi$,
as confirmed in various models.  In the former case the electroweak symmetry remains
unbroken.  What we want is the latter.   In this case $m_{KK} \sim 10 m_W$ and
$m_H \sim 10\,$GeV.  One has too low $m_{KK}$ and too small $m_H$.  

There are two ways to circumvent these difficulties.  One way is to arrange the
matter content such that small $\theta_H$ is obtained.  This is possible as
discussed by many authors, but requires either many additional fields in 
higher dimensional representations in $\cG$, or fine-tuned cancellations
among contributions from various fields.\cite{HHKY, Haba, Csaki2}  
Another way is to consider warped (curved) spacetime 
in extra-dimensions.\mycite{HM}{HS2}$^,$\mycite{Pomarol2}{Contino1}
Astonishingly
the warped spacetime resolves the above problems quite
naturally as discussed below.

\section{$SO(5) \times U(1)$ unification in warped spacetime}

An attractive model is obtained by considering gauge theory
in the Randall-Sundrum (RS) warped spacetime\cite{RS1, GP, Chang} 
whose metric is given by 
\beeq
ds^2 = e^{-2k\sigma(y)} \eta_{\mu\nu} dx^\mu dx^\nu + dy^2
\label{metric1}
\eneq
where $\sigma(y+2\pi R) = \sigma(y) = \sigma(-y)$ and $\sigma(y) = k |y|$
for $|y| \le \pi R$.  The topology of the spacetime is the same as $M^4 \times (S^1/Z_2)$.
The spacetime is an orbifold, with fixed points (branes) at $y=0$ and $\pi R$.  
It has a negative cosmological constant $\Lambda = -k^2$ in the bulk 
five-dimensional spacetime.  The RS spacetime is an anti-de Sitter space sandwiched
by the Planck brane at $y=0$ and the TeV brane at $y=\pi R$.  At low energies
the spacetime resembles four-dimensional Minkowski spacetime.

We consider $SO(5) \times U(1)_{B-L}$ gauge theory\cite{Agashe2}  with gauge couplings $g_A$ and 
$g_B$  defined in the five-dimensional spacetime (\ref{metric1}).  We suppose that 
the structure of the spacetime is determined by physics at the Planck scale and
therefore $k = O(M_\Pl)$.  With the warp factor $e^{-k\pi R}$  the electroweak scale
$m_W$  is naturally generated from the Planck scale. 

The orbifold boundary conditions for the $SO(5)$ and $U(1)_{B-L}$ gauge fields, 
$A_M$ and $B_M$, are given by
$P_0=P_1= {\rm diag} (-1,-1,-1,-1,1)$ and $P_0=P_1=1$ in (\ref{BC1}), respectively.  
With this parity assignment 
the bulk $SO(5) \times U(1)_{B-L}$ symmetry breaks down to 
$SO(4) \times U(1)_{B-L}= SU(2)_L \times SU(2)_R \times U(1)_{B-L}$
 on the branes.  We further break the symmetry on the 
Planck brane by imposing the Dirichlet condition on $A_\mu^{1_R}$, 
$A_\mu^{2_R}$,  and $A_\mu^{\prime 3_R}$ which are even under parity.
Here  $A_\mu^{a_R}$ ($a=1,2,3$) are $SU(2)_R$ gauge fields and  
\beeq
A_\mu^{\prime 3_R} = \frac{g_A  A_\mu^{3_R} - g_B B_\mu}{\sqrt{g_A^2 + g_B^2}}  
~~,~~
A_\mu^Y  =  \frac{g_B  A_\mu^{3_R} + g_A B_\mu}{\sqrt{g_A^2 + g_B^2}} ~~.
\label{newA}
\eneq
$A_\mu^Y $ obeys the Neumann condition on both branes.  As a result the residual 
symmetry is $SU(2)_L \times U(1)_Y$.  The change of the boundary conditions
from Neumann to Dirichlet for $A_\mu^{1_R}$,  $A_\mu^{2_R}$,  and 
$A_\mu^{\prime 3_R}$ is induced by additional dynamics on the Planck brane,
and is consistent with the large gauge invariance.\cite{SH1, HS2, Sakai}

\subsection{Mass spectrum}

There  is one mass scale in the theory, namely $k = O(M_\Pl)$, 
and a few dimensionless parameters $k \pi R$, $g_A/\sqrt{\pi R}$ and
$g_B/\sqrt{\pi R}$.  The Kaluza-Klein mass scale in the RS spacetime is
\beeq
m_{KK} = \frac{\pi k}{e^{k\pi R} - 1} \sim 
\begin{cases}
\pi k e^{-k\pi R} &{\rm for~}  e^{k\pi R} \gg 1 ~, \cr
1/R &{\rm for~}   k \go 0 ~.
\end{cases}
\label{KKscale1}
\eneq

For $\theta_H \not= 0$, $m_W$ and $m_Z$ are given by
\beqn
&&\hskip -1cm
m_W \sim \sqrt{ \frac{k}{\pi R}} ~ e^{-k\pi R} ~ \sin \theta_H \cr
\noalign{\kern 5pt}
&&\hskip -1cm 
m_Z \sim \frac{m_W}{\cos \theta_W} ~~,~~
\sin \theta_W = \frac{g_Y}{\sqrt{g_A^2 + g_Y^2}}
= \frac{g_B}{\sqrt{g_A^2 + 2 g_B^2}} ~~.
\label{WZmass1}
\eeqn
In a generic situation one has $\sin \theta_H = O(1)$.  It follows from the 
relation for $m_W$ that the dimensionless parameter $k\pi R \sim 37$
for $k= O(M_\Pl)$.  

Further (\ref{KKscale1}) and (\ref{WZmass1}) imply that
\beeq
m_{KK} \sim \frac{\pi}{\sin \theta_H}  \sqrt{k \pi R} ~ m_W  ~~.
\label{KKscale2}
\eneq
For moderate values $\theta_H = (0.2 \sim 0.5)\pi$, the Kaluza-Klein scale
turns out $m_{KK} = 2.6 \, {\rm TeV} \sim 1.5 \, {\rm TeV}$, which is
large enough to be consistent with the current experimental limit. 
One of the problems in the gauge-Higgs unification scenario in flat spacetime
mentioned earlier is solved.   In the  Randall-Sundrum spacetime there 
appears an enhancement factor $\sqrt{k\pi R}$.

\begin{figure}[b,h,t]
\centering  \leavevmode
\includegraphics[width=10cm]{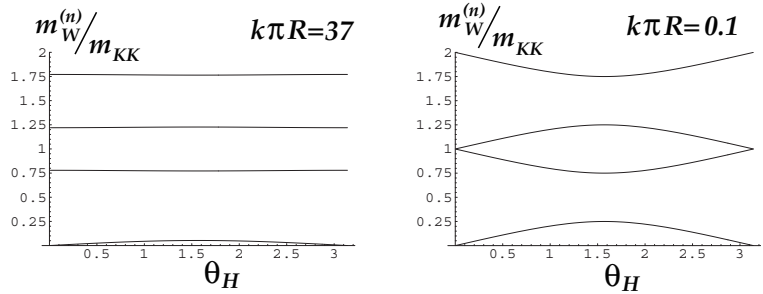}
\hskip .2cm
\caption{$m^{(n)}_W/m_{KK}$ $(n=0,1,2,3)$ 
is plotted for $k\pi R =37$ and 0.1, where
the former corresponds to the realistic case, whereas the latter
is close to the flat spacetime limit ($k\pi R = 0$).  $m_W = m_W^{(0)}$.}
\label{Wmass-fig}
\end{figure}

The mass scale of low energy modes becomes much smaller than the 
Kaluza-Klein mass scale in the warped spacetime.  This can be most clearly
seen by examining the mass spectrum as a function of $\theta_H$ with
various values of $k\pi R$.  See fig.\ \ref{Wmass-fig}.  $m_W/m_{KK}$ has
weak dependence on $\theta_H$ for $k\pi R = 37$ and is much smaller than 1.
In the flat spacetime $m_W/m_{KK}$ becomes $O(0.1)$ for 
$0.1 \pi < \theta_H < 0.9\pi$.

\subsection{Higgs mass and self-couplings}

The Higgs mass and self-couplings are generated by quantum effects, or by radiative 
corrections.   The 4D Higgs field $\vphi_H(x)$ corresponds to four-dimensional
fluctuations of  $\theta_H$.  In the $SO(5) \times U(1)_{B-L}$ model
\beeq
A_y^{\hat 4} (x,y) = \frac{2 \sqrt{2} ~ k ~ e^{2ky} }{g_A ( z_\pi^2-1)}
\bigg\{ \theta_H + 
\frac{g_A}{2} \sqrt{ \frac{ z_\pi^2 -1}{k} } ~ \vphi_H(x) \bigg\}
+\cdots
\label{HiggsField1}
\eneq
where $z_\pi = e^{k\pi R}$.  Thus, the Higgs mass $m_H$, for instance, is 
evaluated from the curvature of $V_\eff(\theta_H)$ at the minimum.  Notice that
$\theta_H$ and $\vphi_H(x)$ appear in the effective potential  
in the combination of
\beeq
\begin{cases}
 \theta_H + \myfrac{\pi g}{\sqrt{2} m_{KK}} \sqrt{\myfrac{k\pi R}{2}} ~ \vphi_H(x)
&{\rm for~}  e^{k\pi R} \gg 1 ~, \cr
 \theta_H + \myfrac{\pi g}{\sqrt{2} m_{KK}}~ \vphi_H(x)
&{\rm for~}   k \go 0 ~,
\end{cases}
\label{HiggsField2}
\eneq
where the 4D $SU(2)_L$ coupling $g$ is given by $g=g_A/\sqrt{\pi R}$.
We observe that $k\pi R/2 \sim 18$ gives various quantities in the warped space 
an enhancement factor compared with those in flat space.

On  general ground the  effective potential at  one loop is
estimated as
\beeq
V_\eff(\theta_H) \sim \frac{3}{128 \pi^6} \, m_{KK}^4  \, f(\theta_H)
\label{effV2}
\eneq
where $f(\theta_H) = O(1)$ in minimal models. The mass  $m_H$ 
and the quartic coupling $\lambda$ (in $\lambda \vphi_H^4/4!$) are evaluated as
\beeq
m_H \sim \sqrt{ \frac{3 \alpha_W}{32 \pi}  f^{(2)}(\theta_H) }
~ \frac{k\pi R}{2}  ~ \frac{m_W}{\sin \theta_H} ~,~~
\lambda \sim  \frac{3 \alpha_W^2}{32} f^{(4)} (\theta_H) 
\bigg(  \frac{k\pi R}{2} \bigg)^2  ~,
\label{HiggsMass2}
\eneq
where $\alpha_W = g^2/4\pi$.
There is ambiguity in $f^{(2)}, f^{(4)}$ which somewhat depend on
detailed content of the model.
Inserting typical values $f^{(2)}, f^{(4)} \sim 4$ and
$\theta_H = (0.1 \sim 0.5) \pi$, one finds that
$m_H = (90 \sim 290) \,$GeV and $\lambda \sim 0.1$.  
Although the precise form of $f(\theta_H)$ depends on  details of the 
model, the feature of the enhancement by the factor $k\pi R/2$ in the 
RS spacetime is general.
The problem of  too small $m_H$ in flat spacetime has been solved.

\subsection{$WWZ$ coupling}

When $\theta_H=0$, the electroweak symmetry $SU(2)_L \times U(1)_Y$
remains unbroken.  The $SU(2)_L$ gauge coupling in four dimensions 
is given by $g=g_A/\sqrt{\pi R}$.  All couplings associated with $W$ and 
$Z$ are  determined  by the  $SU(2)_L \times U(1)_Y$ gauge principle.  
When $\theta_H \not= 0$, things are not so simple in the gauge-Higgs 
unification scenario.

With $\theta_H \not= 0$, $SU(2)_L \times U(1)_Y$ breaks down to $U(1)_{EM}$.
In the standard model the $W$ boson resides in the $SU(2)_L$ group.  
In the $SO(5) \times U(1)_{B-L}$ gauge-Higgs unification model, 
$\theta_H \not= 0$ mixes various components of $SU(2)_L$, $SU(2)_R$ and
$SO(5)/SO(4)$.  It also mixes various Kaluza-Klein excited states.
The eigenstate $W$ and its wave function are determined by 
complete diagonalization.    This poses an interesting question whether or not
the $WWZ$ coupling $g_{WWZ}$, for instance, remains universal as in the standard model.
There is no guarantee for that.

This is an important issue as the LEP2 data on the $W$ pair production rate
agrees with the $WWZ$ coupling in the standard model within
an error of a few percents.    In Table \ref{WWZ-table} the ratio of $g_{WWZ}$
in the gauge-Higgs unification to that in the standard model  is tabulated
for various $\theta_H$ and $k\pi R$.  One sees that for the realistic case 
$k\pi R \sim 35$,  deviation from the standard model is tiny for any values of 
$\theta_H$, whereas in the flat spacetime limit ($k\pi R=0$) substantial deviation
appears for moderate values of $\theta_H$.

\begin{table}[h,t,b]
\tbl{The   ratio of $g_{WWZ}$ in the gauge-Higgs unification 
to that in the standard model 
 for $\theta_H=\pi/10,\pi/4,\pi/2$   and $k\pi R=35,3.5,0.35$. }
{\begin{tabular}{|c|c|c|c|}
 \hline   \rule[-1.8mm]{0mm}{5mm}
 & $\theta_H=\frac{1}{10} \pi$ & $\frac{1}{4} \pi$ & $\frac{1}{2}\pi$  \\ 
 \hline  \rule[-1.8mm]{0mm}{5mm}
 $k\pi R=35$ & ~0.9999987~ & ~0.999964~ & ~0.99985~ \\ 
 \hline  \rule[-1.8mm]{0mm}{5mm}
 3.5 & 0.9999078 & 0.996993 & 0.98460 \\ 
 \hline  \rule[-1.8mm]{0mm}{5mm}
 0.35 & 0.9994990 & 0.979458  & 0.83378 \\ \hline 
 \end{tabular}  }
\label{WWZ-table}
\end{table}

The $WWZ$ coupling remains almost universal in the warped space.
The gauge-Higgs unification scenario in the warped space is consistent 
with the LEP2 data, whereas the scenario in flat space conflicts with the
data unless $\theta_H$ is sufficiently small.

\subsection{$WWH$ and $ZZH$ couplings}

There emerges significant deviation from the standard model in various
couplings of the Higgs boson.  Unlike 4D gauge bosons the 4D Higgs boson
is mostly localized near the TeV brane so that the behavior of wave functions
of various fields on the TeV brane becomes relevant for their couplings to 
the Higgs boson.  

Robust prediction is obtained for the $WWH$ and $ZZH$ couplings
\beeq
 \lambda_{WWH} ~ H \,  W^{\mu\,\dagger}W_\mu 
 +  \half  \lambda_{ZZH} ~ H \, Z^\mu Z_\mu ~.
 \label{WWH1}
\eneq
The detailed matter content affects the effective potential $V_\eff (\theta_H)$,
but the couplings $\lambda_{WWH}$ and $\lambda_{ZZH}$ are determined
independent of such details once $\theta_H$ is given.  One finds that
\beeq
\lambda_{WWH} 
\simeq   gm_W \cdot p_{\rm H}  |\cos\theta_H| ~~,~~ 
 \lambda_{ZZH}   \simeq  
\frac{gm_Z}{\cos \theta_W }\cdot p_{\rm H} |\cos\theta_H| 
 \label{WWH2}
\eneq
where $p_{\rm H} \equiv {\rm sgn}  (\tan\theta_H)$.  Compared with 
the values in the standard model, both couplings are suppressed by a
factor $\cos\theta_H$.  This result can be used to experimentally test the
gauge-Higgs unification scenario.

\subsection{Yukawa coupling}

Couplings of the Higgs boson to quarks and leptons, Yukawa couplings, 
are also subject to nontrivial $\theta_H$-dependent suppression.
The Lagrangian for fermions is given by\cite{GP, Chang}
\beqn
&&\hskip -1cm
i  \psibar \Gamma^a {e_a}^M  \Big\{
\dd_M + \frac{1}{8} \omega_{bcM} [ \Gamma^b, \Gamma^c] - i g_A A_M
 - i \frac{g_B}{2} {\cal Q}_{\rm B-L} B_M \Big\} \psi \cr
\noalign{\kern 5pt}
&&\hskip 1cm
- c\, k \,  \ep(y) \,  \psibar \psi + \hbox{brane interactions} ~.
\label{fermion1}
\eeqn
${\cal Q}_{\rm B-L}$ is a charge of $U(1)_{B-L}$.  
The kink mass term $c k   \ep(y)$  naturally arises in the 
Randall-Sundrum spacetime where  a dimensionless parameter  $c$ for
each fermion multiplet plays a crucial role for determining its wave function.  
There can be ``brane interactions'' between $\psi$ and additional brane fermion
fields defined on one of the branes.  

The Higgs coupling to $\psi$ is contained in the gauge interaction involving
$A_y$.  Non-vanishing $\theta_H$ ($\la A_y \ra \not= 0$)  induces 
a finite fermion mass.  
Although the gauge interaction is universal, the resulting 4D mass and 
Yukawa interaction depend on the wave function in the fifth dimension,
or on $c$ and the brane interactions. This gives flavor-dependent masses
and Yukawa couplings.  In the absence of brane interactions, $c=\pm \onehalf$
gives a fermion a mass of $O(m_W)$.  Light fermions ($e, \mu, \tau, u,d, s, c, b$)
corresponds to $c = (0.6 \sim 0.8)$, whereas a heavy fermion ($t$) to $c \sim 0.4$.
The large hierarchy in the fermion mass spectrum is  explained
by plain distribution in the parameter $c$.

In the minimal standard model the Yukawa coupling  is proportional to the mass
of a fermion.   In the gauge-Higgs unification scenario this relation is 
modified.  In the absence of brane interactions the Yukawa coupling in the 
gauge-Higgs unification in the RS spacetime is suppressed by a factor
$\cos \theta_H$ or $\cos \onehalf \theta_H$ compared with the value
in the standard model.  To realize the observed spectrum of quarks and leptons,
however, one needs to include brane interactions, which in turn affects
the relationship between the mass and Yukawa coupling.  Although 
the relationship depends on details of the model, it is expected that 
it deviates from that in the standard model.

\subsection{Gauge couplings of fermions}

Couplings of quarks and leptons to $W$ and $Z$ also suffer from 
modification, but the amount of deviation from the standard model
turns out tiny.  The $\mu$-$e$ universality in weak interactions
played an important role in the development of the  theory.  In the 
modern language it says that all left-handed leptons and quarks have
the same coupling to the $W$ boson.  It is dictated by the $SU(2)_L$
gauge invariance in four dimensions.  In the gauge-Higgs unification,
however, the universality is not guaranteed at $\theta_H \not= 0$.  
As explained earlier, non-vanishing $\theta_H$ mixes various components
in the gauge group and various levels in the Kaluza-Klein tower.   This 
mixing for fermions depends on, say, the kink mass parameter $c$, and
therefore is not universal.  

For $c > 0.6$ wave functions are mostly localized near the Planck brane
at $y=0$ so that the 4D gauge coupling to $W$ becomes almost
universal for any values of $\theta_H$.  
Define $r_\mu (\theta_H) = g^W_\mu (\theta_H)/g^W_e (\theta_H) -1$
where $g^W_e$ and $g^W_\mu$ are the gauge ($W$) couplings of $e$ and 
$\mu$, respectively.  One finds typically  that $r_\mu \sim - 10^{-8}$ for 
$\theta_H = 0.5 \pi$.   For $\tau$, $r_\tau \sim - 2 \times 10^{-6}$.
These numbers are well within the experimental limit,  being very hard
to test in the near future.  For top quarks,  the deviation becomes bigger
($r_t (0.5 \pi) \sim - 2 \times 10^{-2}$),  but is difficult to measure
accurately.

\section{Flat v.s. Warped}

Why do we need the warped spacetime rather than flat spacetime?
The Randall-Sundrum warped spacetime was originally introduced 
to naturally explain the hierarchy between the Planck scale and weak
scale.  When applied to the gauge-Higgs unification, there are more
benefits.  

See Table \ref{flat-warped-table}.   Both  Higgs mass and Kaluza-Klein
mass scale turn out too small in flat space for moderate values of $\theta_H$.
The $\rho$ parameter  deviates from 1 even at the tree level 
and the $WWZ$ coupling deviates from the value in the standard model 
in falt space.  All these problems are resolved in the Randall-Sundrum
warped space.  Besides the observed fermion spectrum can be 
explained without any fine tuning of the parameters.

All of them indicate that having the Randall-Sundrum warped spacetime 
as background  is not just an accident, but  have a deeper reason.  In this 
regard the holographic interpretation of the model in the AdS/CFT 
correspondence is very suggestive as explored by many authors.

\section{Conclusion}
The prospect of the gauge-Higgs unification in the warped spacetime  is bright.
The Higgs field is identified with the Yang-Mills AB phase in the extra
dimension.   It gives definitive prediction in the Higgs couplings, 
which can be tested at LHC and ILC.    The model has not been 
completed yet.   The most urgent task includes to pin down additional brane 
interactions for fermions so that the observed quark-lepton mass spectrum 
and the CKM and MNS mixing matrices are reproduced.

\begin{table}[t,b]
\tbl{Comparison of the gauge-Higgs unification in the $SO(5) \times U(1)$ 
model in the flat spacetime
$M^4 \times (S^1/Z_2)$ and in the Randall-Sundrum warped spacetime.
$\theta_H = (0.1 \sim 0.5) \pi$.   $k\pi R=37$ for the RS spacetime.
The estimate of $m_H$ has ambiguity in $f^{(2)}(\theta_H)$ 
in (\ref{HiggsMass2}). }
{\begin{tabular}{|c|c|c|}
\hline \rule[-2.2mm]{0mm}{5.5mm} 
Background spacetime &  
 $M^4 \times (S^1/Z_2)$ & ~ Randall-Sundrum ~ \\ 
 \hline  \rule[-2.2mm]{0mm}{5.5mm}
Higgs mass $m_H$
&$3 \sim 16 \,$GeV &$100 \sim 300 \,$GeV  \\ 
 \hline  \rule[-2.2mm]{0mm}{5.5mm}
KK  mass scale $m_{KK}$
&$0.3 \sim 1.1 \,$TeV &$1.5 \sim 5.0 \,$TeV  \\ 
\hline  \rule[-2.2mm]{0mm}{5.5mm}
$\sin \theta_W$, $\rho$ & deviation from SM &OK  \\ 
\hline  \rule[-2.2mm]{0mm}{5.5mm}
$WWZ$ coupling  & deviation from SM &OK (almost universal) \\ 
\hline  \rule[-2.2mm]{0mm}{5.5mm}
$WWH$ coupling  & --- &suppressed by $ \cos\theta_H$  \\ 
\hline  \rule[-2.2mm]{0mm}{5.5mm}
$ZZH$ coupling  &--- &suppressed by $ \cos\theta_H$  \\ 
\hline  \rule[-2.2mm]{0mm}{5.5mm}
Quark-lepton  spectrum
&fine tuning necessary 
&natural hierarchy    \\ 
\hline  \rule[-2.2mm]{0mm}{5.5mm}
Yukawa couplings  &--- &generally suppressed   \\ 
\hline 
 \end{tabular}}
\label{flat-warped-table}
\end{table}

\section*{Acknowledgements}

This work was supported in part by  Scientific Grants from the Ministry of 
Education and Science, Grant No.\ 17540257,
Grant No.\ 13135215 and Grant No.\ 18204024.  The author would like to thank
the Aspen Center for Physics for its hospitality where a part of this work 
was performed.

\def\jnl#1#2#3#4{{#1}{\bf #2} (#4) #3}

\def\reftitle#1{}                

\def\Zphys{{\em Z.\ Phys.} }
\def\jssc{{\em J.\ Solid State Chem.\ }}
\def\jpsJ{{\em J.\ Phys.\ Soc.\ Japan }}
\def\ptps{{\em Prog.\ Theoret.\ Phys.\ Suppl.\ }}
\def\PTP{{\em Prog.\ Theoret.\ Phys.\  }}

\def\JMP{{\em J. Math.\ Phys.} }
\def\NPB{{\em Nucl.\ Phys.} B}
\def\NP{{\em Nucl.\ Phys.} }
\def\PLB{{\em Phys.\ Lett.} B}
\def\PL{{\em Phys.\ Lett.} }
\def\PRL{\em Phys.\ Rev.\ Lett. }
\def\PRB{{\em Phys.\ Rev.} B}
\def\PRD{{\em Phys.\ Rev.} D}
\def\PRe{{\em Phys.\ Rep.} }
\def\AP{{\em Ann.\ Phys.\ (N.Y.)} }
\def\RMP{{\em Rev.\ Mod.\ Phys.} }
\def\ZPC{{\em Z.\ Phys.} C}
\def\SCI{\em Science}
\def\CMP{\em Comm.\ Math.\ Phys. }
\def\MPLA{{\em Mod.\ Phys.\ Lett.} A}
\def\IJMPA{{\em Int.\ J.\ Mod.\ Phys.} A}
\def\IJMPB{{\em Int.\ J.\ Mod.\ Phys.} B}
\def\EPJC{{\em Eur.\ Phys.\ J.} C}
\def\PR{{\em Phys.\ Rev.} }
\def\JHEP{{\em JHEP} }
\def\cmp{{\em Com.\ Math.\ Phys.}}
\def\JPA{{\em J.\  Phys.} A}
\def\JPG{{\em J.\  Phys.} G}
\def\NJP{{\em New.\ J.\  Phys.} }
\def\CQG{\em Class.\ Quant.\ Grav. }
\def\ATMP{{\em Adv.\ Theoret.\ Math.\ Phys.} }
\def\ibid{{\em ibid.} }


\end{document}